# Work Function Characterization of Directionally Solidified LaB$_6$–VB$_2$ Eutectic

Tyson C. Back[1,2], Steven B. Fairchild[2], John J. Boeckl[2], Marc Cahay[3], Floor Derkink[4,5], Gong Chen[4], Andreas K. Schmid[4], Ali Sayir[6]

[1] University of Dayton Research Institute, 300 College Park, Dayton, OH 45469-0170, USA
[2] Air Force Research Laboratory, Materials and Manufacturing Directorate, 3005 Hobson Way, Wright-Patterson Air Force Base, OH 45433, USA
[3] Spintronics and Vacuum Nanoelectronics Laboratory, University of Cincinnati, Cincinnati, Ohio 45221, USA
[4] Physics of Interfaces and Nanomaterials, MESA+ Institute for Nanotechnology, University of Twente, P.O. Box 217, 7500 AE Enschede, The Netherlands
[5] NCEM, Molecular Foundry, Lawrence Berkeley National Laboratory, Berkeley, CA 94720, USA
[6] NASA Glenn Research Center, Cleveland, OH 44135, USA

**ABSTRACT**

With its low work function and high mechanical strength, the LaB$_6$/VB$_2$ eutectic system is an interesting candidate for high performance thermionic emitters. For the development of device applications, it is important to understand the origin, value, and spatial distribution of the work function in this system. Here we combine thermal emission electron microscopy and low energy electron microscopy with Auger electron spectroscopy and physical vapour deposition of the constituent elements to explore physical and chemical conditions governing the work function of these surfaces. Our results include the observation that work function is lower (and emission intensity is higher) on VB$_2$ inclusions than on the LaB$_6$ matrix. We also observe that the deposition of atomic monolayer doses of vanadium results in surprisingly significant lowering of the work function with values as low as 1.1 eV.

**Introduction**

LaB$_6$ has been used as thermionic emitter for several decades. It was recognized early on that LaB$_6$ had many properties that were beneficial for use as a cathode, such as low work function (2.7 eV) and lower operating temperatures (~1500 K) [1]. LaB$_6$ also has an energy spread roughly half that of tungsten under the same accelerating voltage [2]. All of which are an improvement over the standard tungsten electron sources with the main limiting factor to more widespread use being cost. As a result, LaB$_6$ has been employed in a wide variety of applications requiring an electron source, which include electron microscopes, travelling wave tubes, and Hall/ion thrusters.

Directionally solidified eutectics consist of a two-phase material with one phase distributed throughout the matrix of the second phase. This class of materials have been shown to have desirable high-temperature mechanical properties compared to existing composites [3-9]. Somewhat more recently LaB$_6$ directionally solidified eutectic (DSE) materials have been shown to offer further mechanical improvements over standard LaB$_6$ [10]. LaB$_6$ DSEs consist of a LaB$_6$ matrix phase with a transition metal di-boride phase that forms fibers homogenously throughout the matrix. Typical transition metals consist of Zr, Hf, Ti, and V. The improvement in mechanical properties is attributed to the interface between the two phases in the eutectic. Additionally, this material system has also shown significant improvements in thermionic emission current density when compared with standard single crystal LaB$_6$ [11, 12]. LaB$_6$/VB$_2$ was shown to have an order of magnitude improvement in current density when compared to single crystal LaB$_6$ [12].

The combination of high mechanical strength and current density make LaB$_6$ DSEs ideal candidates for use in applications where high power and long term stability are critical such as use in Hall/ion thrusters where cathode requirements up to 400 A and $10^4$ hours could be necessary [13]. Although previous electron emission studies of LaB6 DSEs have shown significant improvements over existing thermionic cathodes, a fundamental understanding of why is lacking. It is the purpose of this paper to investigate dynamic changes in work function for LaB$_6$/VB$_2$ with low energy electron microscopy (LEEM). The thermionic electron emission microscopy (ThEEM) imaging mode and reflectivity curves were used to characterize the work function of the surface under stoichiometric and nonstoichiometric conditions.

**Experimental Procedures**

The $LaB_6/VB_2$ samples were directionally solidified by a zone melting technique previously described elsewhere [10]. The crystal growth resulted in cylindrical rods which were then cut and polished. After polishing the samples were transferred into a spin polarized low energy electron microscope (SPLEEM) at the National Center for Electron Microscopy at Lawrence Berkeley National Laboratory. Before imaging the samples were introduced into a sample preparation chamber for cleaning that consisted of Ar ion sputtering in a background of $O_2$, 3.0 x $10^{-8}$ Torr. During sputtering the sample was flashed to 1250 $^0$C for 40s with a final flash in vacuum. This was repeated multiple times until the sample surface was free of carbon and oxygen which was checked with Auger electron spectroscopy (AES). Once clean, the sample was transferred into the SPLEEM imaging chamber which maintained a base pressure of 2 x $10^{-11}$ Torr. All images were acquired in the bright field imaging mode. The SPLEEM setup has been described elsewhere [14, 15].

In order to determine work function, a series of images were acquired by systematically changing the starting voltage on the sample to generate a reflectivity curve. The curves can be used to determine at what point the transition from mirror mode to scattering mode, sometimes referred to as the MEM-LEEM transition, occurred. This transition can be used to determine the work function of the sample surface using the relationship $\phi_S = eV_{onset} + \phi_G$, where $\phi_S$ is the work function of the sample in eV, $V_{onset}$ is the threshold voltage in volts and $\phi_G$ is the effective work function of the electron gun [16-19]. In order to accurately determine the work function of the sample by this method the effective work function of the electron gun in the LEEM must be known. This was calibrated against a well-known W(110) surface which has a work function of 5.2 eV[20]. The series of images created by changing the starting voltage on the sample were used to generate a reflectivity curve for each pixel in the image. These curves were used to create a spatially resolved work function map. In order to characterize the effects of non-stoichiometric surfaces on work function the samples were dosed with monolayer (ML) coverages of La, V, and B separately during image acquisition. V and B were deposited by *e*-beam evaporation and La by thermal evaporation. For calibration of deposition rate each individual elemental component was deposited onto a clean $LaB_6$ surface. The deposition rate was determined by monitoring the image intensity oscillations that are consistent with atomic layer-by-layer growth. Samples were also imaged in thermal emission imaging mode or ThEEM. In ThEEM only thermally emitted electrons from the sample are used for imaging. In this imaging mode, thermionic emission curves were obtained by plotting the total intensity of the image as a function of temperature [21].

**Results and Discussion**

It was previously shown that $LaB_6$-$MeB_2$ (where Me = V, Zr, Ti, and Hf) materials exhibit significant improvements in thermal emission current density compared to pure $LaB_6$ [12]. It was found that for all the compositions tested it was the eutectic composition that always yielded the highest current density[22]. Of the transition metal di-borides tested, the $VB_2$ compound yielded the highest current density. Fig. 1 shows a LEEM image of the $LaB_6$/$VB_2$ surface. The dark areas in the image consist of circular features roughly 500 nm in diameter which are the $VB_2$ phase. The lighter areas in the image are the $LaB_6$ matrix phase.

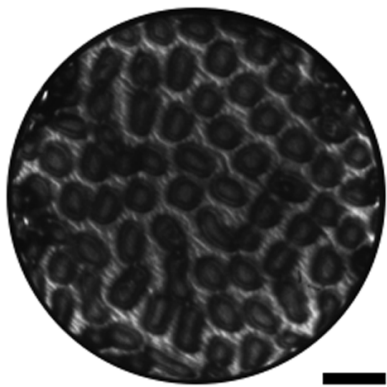

Figure 1 LEEM image of $LaB_6$/$VB_2$ surface, electron landing energy 2 eV. The eutectic consist of a $LaB_6$ matrix with $VB_2$ rods. The rods are typically ~600 nm in diameter. Scale bar is 1µm.

Taran *et al.* conjectured that the origin for the improvement in electron emission was due to improved La diffusion mobility along the interface between the two phases compared to bulk $LaB_6$. This process resulted in excess La concentration at the surface of the cathode. Berger *et al.* recently showed that for the $LaB_6$-$ZrB_2$ eutectic enhanced emission around the phase boundaries was evident in ThEEM[23]. Diffusion at the phase boundaries is a likely mechanism for the observed emission enhancement. It was shown that by replacing $ZrB_2$ with a solid solution of $(Zr, Ti)B_2$ the emission activity decreased [11]. This was partially attributed to the presence of Ti atoms at the phase boundary interface. The solid solution di-boride is thought to form a more perfect interface, limiting the diffusion of La. Fig. 2(a,b) show a LEEM image of the $LaB_6$-$VB_2$ surface with corresponding work function map. It can be seen in Fig 2(b) that low work function areas are primarily concentrated around the phase boundary between the two materials, with values ~1.6 eV. It should be noted that the surface cleaning procedure involves multiple high temperature flashes to 1250 $^0$C. It is possible that the cleaning procedure promoted diffusion of La to the surface of the eutectic similar to an effect previously shown with $LaB_6$-$ZrB_2$ [23].

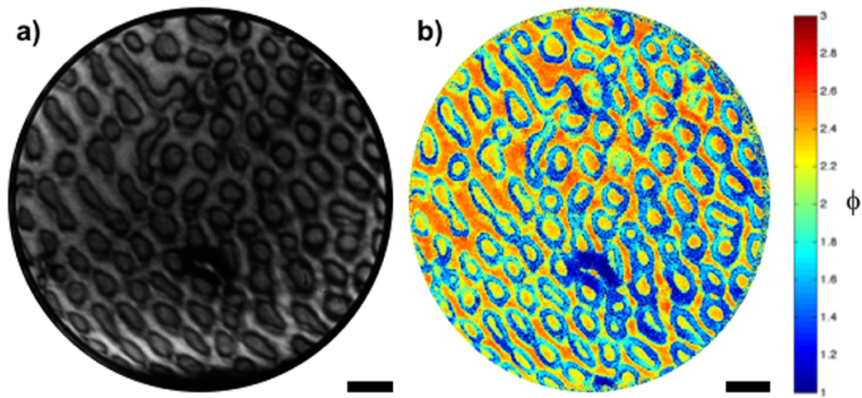

Figure 2 a) LEEM image of $LaB_6$/$VB_2$ surface, electron landing energy 1.1 eV; b) work function map created from a). Most of the low work function areas are concentrated around the phase boundaries between the $LaB_6$ (matrix) and the $VB_2$ (rods). Scale bar is 1μm for both images.

A similar result was obtained with ThEEM. Fig 3(a) shows a ThEEM image at 977 $^0$C. Most of the intensity in the image is concentrated on the $VB_2$ phase. Fig. 3(b) shows a thermionic emission curve generated from image intensity at various temperatures. Using the Richardson-Dushman equation [24], *I*

$= AT^2 exp\text{-}(\phi/kT)$, where $A$ is a material specific constant, $T$ is temperature, $\phi$ is the work function and $k$ is the Boltzmann's constant. In this imaging mode the dominant physical property that contributes to image contrast is work function. Low work function areas will appear brighter. This is somewhat contradictory to the work function map in Fig 2(b), which showed the $VB_2$ phase to have similar work function as the $LaB_6$ matrix. The origin of this discrepancy is unknown at this time. The fact that the work function observed in the map and one calculated with the Richardson-Dushman equation are nearly identical indicates the origin of the work function are possibly the same. Given the large temperature difference, surface diffusion is likely to play a role in the emission activity shift from phase boundaries at room temperature to the primarily the $VB_2$ at 977 $^0C$.

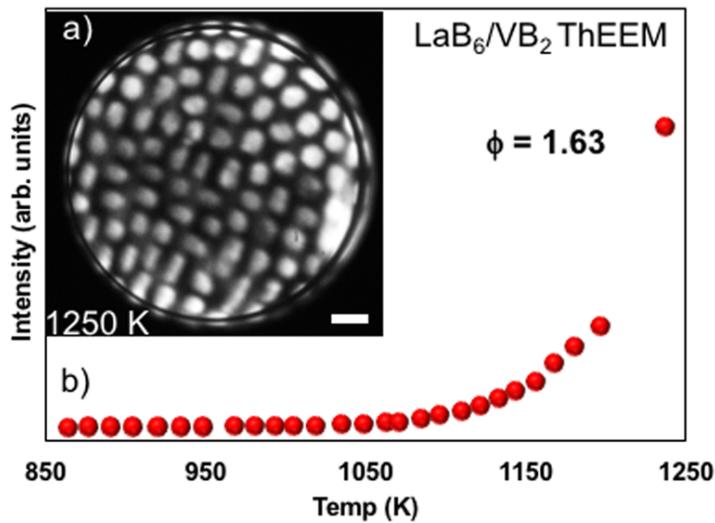

Figure 3 a) ThEEM image of $LaB_6/VB_2$ at 1250 K. Most of the intensity in the image is on or near the $VB_2$ phase. Scale bar is 1μm. b) Thermionic emission curve generated from ThEEM images. A work function of 1.63 eV was calculated using the Richardson-Dushman equation.

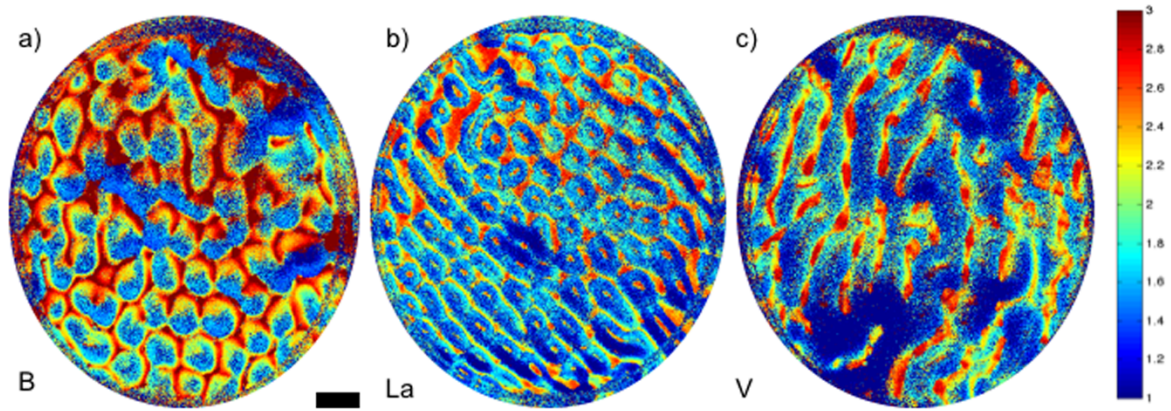

Figure 4 1 ML depositions of a) B, b) La, and c) V. The B deposition showed an overall increase in the work function while both the La and V deposition showed decrease. The scale bar is 1 μm and is applicable to all images.

Although the SPLEEM is capable of heating the samples to temperatures that are typical of a working thermionic emitter, imaging at those conditions for extended periods of time is challenging for many reasons including detrimental *e*-beam heating of the objective lens and drift. Previous work involved thermionic emission experiments that exceeded 100 hours [11]. However, simulating the effects of changes in stoichiometry by dosing the surface is quite easy in the SPLEEM. To do this La, V, and B were dosed separately on the surface, to roughly 1 ML coverage, by *e*-beam and thermal evaporation during image acquisition at room temperature. Fig. 4(a,b,c) shows the results of the dosing experiments. From B dosing results shown in Fig 4a it can be seen that the matrix has a significantly higher work function than the $VB_2$ phase. Higher work functions are also observed around the phase boundaries. This agrees well with the ThEEM image shown in Fig 3a, which showed the primary emission areas to be on the $VB_2$ phase. La dosing shown in Fig 4b yielded low work function areas that were centered around the $VB_2$ phase. The V dosing shown if Fig 4c showed the most significant change in work function. Work functions as low as 1.1 eV were observed on the surface.

Qualitatively, the work function maps in Fig. 4 indicate an overall increase in the work with excess boron and decrease with excess La and V. It is apparent from the maps that V yielded the lowest work function change with values as low 1.1 eV. This suggest that La diffusion may not play a role in thermionic emission under these experimental conditions. It should be noted that previous thermionic emission experiments were conducted at temperatures much greater than 1000 $^0$C. The work function analysis and thermionic emission imaging presented in this work represents emission characteristics well below that regime.

## Summary

The work function of the $LaB_6$-$VB_2$ DSE was characterized through analysis of reflectivity curves acquired in the LEEM and ThEEM. At room temperature, low work function areas were observed around the phase boundaries. Previous work would suggest that these areas were the result of La diffusion. However, for the first time it was shown through ThEEM and elemental dosing that vanadium may be responsible for the observed low work function areas, as low as 1.1 eV, at least in the low temperature, <1000 $^0$C, regime.

## Acknowledgements

TCB gratefully acknowledges support from U.S. Air Force Office of Scientific Research. Work at the Molecular Foundry was supported by the Office of Science, Office of Basic Energy Sciences, of the U.S. Department of Energy under Contract No. DE-AC02-05CH11231